\documentclass[aps,pra,preprint,groupedaddress]{revtex4}
 
\begin{document}
 
\preprint{LA-UR-07-7672}
 
\title{Monte Carlo Determination of Multiple Extremal Eigenpairs}
 
 
\author{T. E. Booth}
\affiliation{ Applied Physics Division, Los Alamos National Laboratory, Los Alamos, NM 87545}
\author{J. E. Gubernatis}
\affiliation{Theoretical Division, Los Alamos National Laboratory, Los Alamos, NM 87545}

 
\date{\today}

\begin{abstract}
We present a Monte Carlo algorithm that allows the simultaneous determination of a few extremal eigenpairs of a very large matrix without the need to compute the inner product of two vectors or store all the components of any one vector. The new algorithm, a Monte Carlo implementation of a deterministic one we recently benchmarked, is an extension of the power  method. In the implementation presented, we used a basic Monte Carlo splitting and termination method called the comb, incorporated the weight cancellation method of Arnow {\it et al.\/}, and exploited a new sampling method, the sewing method, that does a large state space sampling as a succession of small state space samplings. We illustrate the effectiveness of the algorithm by its determination of the two largest eigenvalues of the transfer matrices for variously-sized two-dimensional, zero field Ising models. While very likely useful for other transfer matrix problems, the algorithm is however quite general and should find application to a larger variety of problems requiring a few dominant eigenvalues of a matrix.

\end{abstract}

 
\maketitle
 
 
\section{Introduction}
A common problem in computational physics is computing the eigenpairs of large matrices. We will present a new Monte Carlo algorithm that allows the simultaneous determination of a few extremal eigenpairs of a very large matrix without the need to orthogonalize pairs of vectors to each other or store all the components of any vector. This algorithm is a Monte Carlo implementation of deterministic one we recently benchmarked \cite{gubernatis}. The newly benchmarked algorithm is based on a refinement of the power method recently developed by Booth \cite{booth1,booth2} and does not require, as does the standard Ritz estimator \cite{wilkinson,golub}, the explicit computation of the inner product of two vectors. This feature is of special importance for Monte Carlo use because Monte Carlo sampling provides only successive estimates of eigenvectors represented by a very small subsets of their possible components. For such a situation, the explicit computation of an inner product is impossible.

The basic power method is the traditional starting point for a Monte Carlo determination of the eigenpair associated with the eigenvalues of largest absolute value $\lambda _1$. While various versions of the Monte Carlo power method often compute this dominant eigenvalue very well, computing subdominant eigenvalues $ \lambda _2 ,\lambda _{3,} \ldots $ has often proven much more difficult and is much less frequently attempted. Our Monte Carlo power method computes multiple extremal eigenpairs simultaneously. The particular algorithm presented uses a basic Monte Carlo splitting and termination technique called the comb \cite{comb1,comb2}, incorporates the weight cancellation method of Arnow {\it et al.} \cite{arnow}, and exploits a new sampling method, the sewing method  \cite{booth08}, that does a large state space sampling as a succession of small state space samplings. 

In refining the power method \cite{gubernatis}, we were targeting its use on matrices so large that they are unassailable deterministically because no single vector can be stored in memory. As the system size increases, finding a few extremal eigenpairs of the transfer matrix of the two-dimensional Ising model becomes such a problem. We will illustrate the effectiveness of the algorithm by determining the two largest eigenvalues of the transfer matrices for variously-sized two-dimensional zero field Ising models, exploiting Onsager's exact results \cite{onsager,thompson} for their values as benchmarks. We comment that  two extremal eigenvalues of this matrix are of significant physical interest: the logarithm of $\lambda _1 $ is proportional to the free energy, and the logarithm of the ratio $ \lambda _2 /\lambda _{1} $ is proportional to the reciprocal of the correlation length that controls long range spin correlations near the critical point. Although our algorithm is extendable to finding more than two extremal eigenpairs, we will focus on finding just $ \lambda _1$ and $\lambda _{2} $ \cite{onsager,thompson}.

In the next section, Section~II, we summarize the basic features of the transfer matrix of the two-dimensional Ising model in a zero magnetic field. In subsequent sections we will reference these features to make our algorithm presentation more concrete. In Section~III, we summarize our extension of the power method for the determination of the two eigenpairs corresponding to the two eigenvalues of largest absolute value. Then, in Section~IV, we discuss the basics of our Monte Carlo implementation of this algorithm. We first use the  algorithm for medium-sized matrices for which computer memory is adequate to store the eigenvectors and then use it for much larger-sized matrices for which it is not. For the Monte Carlo sampling of states in latter case, we use the sewing algorithm \cite{booth08} that facilitates sampling of states from a large space from a smaller space. Results presented are for the determination of the two largest eigenvalues of the Ising transfer matrix for various lattice sizes. In the last section, we summarize our work and comment on its likely application to other systems.

We note that the intent of the present application is presenting and benchmarking a new and relatively  general numerical method and not numerically studying the finite-size scaling of the eigenvalues of the matrix used for the benchmarking. Because we reproduce the exact eigenvalues to satisfactory accuracy, our estimates will enjoy the same scaling as the well known exact results for this problem. When the magnetic field is not zero and the eigenvalues are not exactly known, various researchers have calculated up to the four largest eigenvalues by a deterministic version of the power method and have made extensive studies of the scaling of these eigenvalues \cite{deterministic}. With the algorithm to be presented we have reproduced the first two of these eigenvalues and hence would reproduce the basic scaling. A more extensive study of the eigenvalues of the field dependent Ising model will be presented elsewhere \cite{kamiya}. The Monte Carlo approach increases the lattice sizes accessible. We note that these deterministic calculations required multiple  processors. All our calculations were done on a single processor. 

\section{Ising Model}

In his much celebrated work Onsager calculated many of the properties of the two-dimensional Ising model exactly \cite{onsager}. His calculations started with the expression of the partition function in terms of its transfer matrix \cite{onsager,montroll,camp}. He then found all the eigenvalues of this matrix analytically and showed that the scaling the dominant eigenvalue with the area in the thermodynamic limit (the area of the model approaching infinity) implied the onset of long-range ordering among the spin variables. The ratio of the second largest eigenvalue to the first is associated with the spatial behavior of this ordering. At the critical temperature, this ratio approaches unity as the area approaches infinity.
 
We will consider an $m\times n$ Ising model defined with periodic boundary conditions in one direction and open boundary conditions in the other.  The two-dimensional Ising model's energy is
\begin{equation}
E\left\{ \mu  \right\} =  - J\sum\limits_{i = 1}^{m-1} {\sum\limits_{j = 1}^n {\mu _{i,j} \mu _{i+1,j } } }  -
J\sum\limits_{i = 1}^m {\sum\limits_{j = 1}^n {\mu _{i,j} \mu _{i,j + 1} } }
\label{eq:energy}
\end{equation}
Here, $(i,j)$ are the coordinates of a lattice site. The Ising spin variable $\mu_{i,j}$ on each site has the value of $ \pm 1$, the exchange constant $J>0$,  and $\mu _{i,m + 1}  = \mu _{i,1}$. The symbol 
\begin{equation}
\sigma _j  = \left( {\mu _{1,j} ,\mu _{2,j} , \ldots ,\mu _{m,j} } \right)
\label{eq:state}
\end{equation}
denotes a column configuration of Ising spins and there are $2^m$ possible configurations for each column.
  
The transfer matrix $A(\sigma,\sigma')$  follows from a re-expression of the partition function \cite{thompson}
\begin{eqnarray}
 Z\left( {m,m} \right) & = &\sum\limits_{\left\{ \mu  \right\}} {\exp \left[ { - \nu E\left( {\left\{ \mu  \right\}} \right)} \right]}  \nonumber \\
  & = & \sum\limits_{\sigma _1 , \ldots ,\sigma _m } {A(\sigma _1 ,\sigma _2 )} A(\sigma _2 ,\sigma _3 )
  \cdots A(\sigma _{m - 1} ,\sigma _m )A(\sigma _m ,\sigma _1 ) \nonumber \\
  & = &\sum\limits_{\sigma _1 } {A^n (\sigma _1 ,\sigma _1 )}  
 \end{eqnarray}
where
$\nu=J/k_B T$, $k_B$ is Boltzmann's constant, and $T$ is the temperature, and  $A(\sigma,\sigma')$ is a $2^m \times 2^m$ matrix whose elements are
\begin{equation}
A\left( \sigma ,\sigma'\right) = \exp \left(\nu \sum\limits_{k = 1}^{m -1}
\mu _k \mu _{k + 1} \right)  \exp \left(\nu \sum\limits_{k = 1}^m
\mu _k \mu _k^{'} \right)
\label{eq:transfer_matrix}
\end{equation}
As customary for Ising model simulations, we represent a configuration $\sigma$ by the first $m$ bits of an integer between $0$ and $2^m-1$, with a set bit being a $+1$ Ising spin and an unset bit being a $-1$ spin. This convention maps the matrix element $A(\sigma,\sigma')$ between configurations to an element $A_{ij}$ between integers.

We note that $A(\sigma,\sigma')$ is asymmetric and its elements are greater than zero so the matrix is maximally dense and hence irreducible. Because of the positivity and irreducibility the Perron-Frobenius Theorem \cite{wilf} says that  dominant eigenvalue is real, positive, and non-degenerate and all components of the corresponding eigenstate are real and have the same sign. The two largest eigenvalues of $A$, for {\it finite}  $m$, are \cite{kaufman}
\[
\lambda _{\rm{1}}  = \left( {2\sinh 2\nu } \right)^{m/2} \exp \left[ {\frac{1}{2}\left( {\eta _1  + \eta _3  +  \cdots  + \eta _{2m - 1} } \right)} \right]
\]
\begin{equation}
\lambda _{\rm{2}}  = \left( {2\sinh 2\nu } \right)^{m/2} \exp \left[ {\frac{1}{2}\left( {\eta _2  + \eta _4  +  \cdots  + \eta _{2m} } \right)} \right]
\label{eq:exact}
\end{equation}
where
\begin{equation}
\cosh \eta _k  = \cosh 2\nu \coth 2\nu  - \cos \frac{{\pi k}}{m}
\end{equation}
The transfer matrix of the Ising model can be symmetrized \cite{thompson} but we saw no computational advantage for using this form.

\section{Power Method}

For some real-valued $M\times M$ matrix $A$, not necessarily symmetric, we will be concerned with the $M$ eigenpairs $(\lambda_\alpha,\psi_\alpha)$ satisfying
\begin{equation}
A\psi_\alpha = \lambda _\alpha\psi_\alpha
\label{eq:eigenvalue}
\end{equation}
In the simplest application of the power method \cite{wilkinson}, an iteration is started with some  $\psi$, normalized in a convenient, but otherwise relatively arbitrary, manner and consists of iterating the two steps
\begin{equation}
\begin{array}{c}
 \phi = A\psi  \\
 \psi = \phi / \| \phi \| \\
 \end{array}
\label{eq:power_method}
\end{equation}
until some convergence criterion is met. If we write
\begin{equation}
\psi= \sum\limits_{\alpha = 1}^M {\omega_\alpha \psi_\alpha }
\end{equation}
and if $\left| {\lambda _1 } \right| > \left| {\lambda _2 } \right| \ge \left|{\lambda _3 } \right| \ge \cdots  \ge \left| {\lambda _N } \right|$, then after $n$ iterations
\begin{equation}
A^n\psi  = \lambda _1^n \left[ {\omega_1 \psi_1  + \sum\limits_{\alpha= 2}^M {\omega_\alpha \left( {\frac{{\lambda _\alpha}}{{\lambda _1 }}} \right)^n \psi_\alpha} } \right]
\label{eq:pm}
\end{equation}
Accordingly, as $n\rightarrow \infty$,
\begin{eqnarray}
 \psi  &\to& \psi_1 /\|\psi_1\| \nonumber \\
 \|\phi\| &\to& \lambda _1 
 \end{eqnarray}
Thus, the dominant eigenpair is simultaneously determined. For the norm of the vector $\phi$ whose components are $\phi_i$, a frequent choice is
\begin{equation}
\parallel \phi \parallel \equiv \parallel \phi \parallel_\infty = \max_i |\phi_i|
\end{equation}

For deterministic calculations of a few dominant eigenpairs, say $N$, one of two approaches are typically tried. One approach is to use the power method to determine the dominant eigenpair, use deflation to project out this state out of the matrix, and then reuse the power method on the deflated matrix. To determine several eigenpairs simultaneously, the power method can be generalized to 
\begin{equation}
 \Phi  = A\Psi
\end{equation}
where $\Phi$ and $\Psi$ are $M\times N$ matrices whose columns are orthogonalized to each other. This orthogonality needs maintenance throughout the computation or else all $N$ vectors, represented by the columns of the initial $\Psi$, will converge to the one associated with the dominant eigenvalue \cite{golub}. This approach is called orthogonal \cite{golub} or simultaneous iteration \cite{pissanetsky}.

For Monte Carlo calculations of a few dominant eigenpairs, we are proposing a quite different approach based on Booth's proposed refinement of the power method \cite{booth1,booth2}. This refinement  uses the observation that for any eigenpair $(\lambda,\psi)$ and for each non-zero {\it component\/} of the eigenvector, the eigenvalue equation $A\psi  = \lambda\psi$ can be rewritten as
\begin{equation}
\lambda  = \frac{{\sum\limits_j  {A_{i j } \psi _j  } }}{{\psi _i  }}
\label{eq:eigenvalue_estimator}
\end{equation}
and that similar equations can also be written for any number of groupings of components,
\begin{equation}
\lambda  = \frac{{\sum\limits_{i  \in R_1 } {\sum\limits_j  {A_{i j } \psi _j  } } }}{{\sum\limits_{i  \in R_1 } {\psi _i  } }} 
= \frac{{\sum\limits_{i  \in R_2 } {\sum\limits_j  {A_{i j } \psi _j  } } }}{{\sum\limits_{i  \in R_2 } {\psi _i  } }} 
=  \cdots  
= \frac{{\sum\limits_{i  \in R_L } {\sum\limits_j  {A_{i j } \psi _j  } } }}{{\sum\limits_{i  \in R_L } {\psi _i  } }}
\label{eq:groupings}
\end{equation}
where the $R_i$ are rules for different groupings. The groupings are quite flexible: they can overlap  and their union need not cover the entire space. In addition, any two groupings, say 1 and 2, imply
\begin{equation}
\sum\limits_{i  \in R_2 } {\psi _i  } \sum\limits_{i  \in R_1 } {\sum\limits_j  {A_{i j } \psi _j  } }  = \sum\limits_{i  \in R_1 } {\psi _i  } \sum\limits_{i  \in R2} {\sum\limits_j  {A_{i j } \psi _j  } } 
\label{eq:cross_product}
\end{equation}
From $L$ groupings of the components, Booth constructed $L$ estimators for the $L$ largest eigenvalues and forced them to become equal by adjusting certain parameters at each iteration step. For the dominant two eigenvalues, we will do something similar.

First we note that for almost any starting point $\psi = \sum\nolimits_\alpha {\omega_\alpha \psi _\alpha }$, the power method will converge to $ \left( {\lambda _1 ,\psi _1 } \right)$. To find the two extremal eigenvalues, we need two normalized, but not necessarily orthogonal, starting points $\psi'  = \sum\nolimits_\alpha {\omega^\prime_\alpha \psi _\alpha } $ and $\psi'' = \sum\nolimits_\alpha {\omega'' _\alpha \psi _\alpha } $ \cite{gubernatis}.  At each step, we will apply $A$ to them individually. Without any intervention  both will project the same dominant eigenfunction  so at each step we adjust the relationship between their {\it sum} to direct one to the dominant state and the other to the next dominant one.

Formally, we start the iteration with $\psi=\psi'+\eta\psi''$. Suppose at the $n^{th}$ step, $\psi'$ and $\psi''$ have iterated to $\hat\psi'$ and $\hat\psi''$, then at the $(n+1)^{th}$ step we invoke (\ref{eq:groupings}) and 
and (\ref{eq:cross_product}) and require that
\begin{equation}
\frac{{\sum\limits_{i  \in R_1 } {\sum\limits_j  {A_{i j } \hat \psi '_j  } }  + \eta \sum\limits_{i  \in R_1 } {\sum\limits_j  {A_{i j } \hat \psi ''_j  } } }}{{\sum\limits_{i  \in R_1 } {\hat \psi '_i  }  + \eta \sum\limits_{i  \in R_1 } {\hat \psi ''_i  } }} = \frac{{\sum\limits_{i  \in R_2 } {\sum\limits_j  {A_{i j } \hat \psi '_j  } }  + \eta \sum\limits_{i  \in R_2} {\sum\limits_j  {A_{i j } \hat \psi ''_j  } } }}{{\sum\limits_{i  \in R_2 } {\hat \psi '_i  }  + \eta \sum\limits_{i  \in R_2 } {\hat \psi ''_i  } }}
\label{eq:balance}
\end{equation}
which leads to
\begin{equation}
q_2\eta^2+q_1\eta+q_0=0
\label{eq:quadratic}
\end{equation}
with
\begin{eqnarray}
q_2 &=& \sum\limits_{i  \in R_2 } {\hat \psi ''_i  } \sum\limits_{i  \in R_1 } {\sum\limits_j  {A_{i j } \hat \psi ''_j  } }  - \sum\limits_{i  \in R_1 } {\hat \psi ''_i  } \sum\limits_{i  \in R_2} {\sum\limits_j  {A_{i j } \hat \psi ''_j  } } \nonumber \\
q_1 &=& \sum\limits_{i  \in R_2 } {\hat \psi ''_i  } \sum\limits_{i  \in R_1 } {\sum\limits_j  {A_{i j } \hat \psi '_j  } }  - \sum\limits_{i  \in R_1 } {\hat \psi ''_i  } \sum\limits_{i  \in R_2 } {\sum\limits_j  {A_{i j } \hat \psi '_j  } } \nonumber \\
    &+&  \sum\limits_{i  \in R_2 } {\hat \psi '_i  } \sum\limits_{i  \in R_1 } {\sum\limits_j  {A_{i j } \hat \psi ''_j  } }  - \sum\limits_{i  \in R_1 } {\hat \psi '_i  } \sum\limits_{i  \in R_2 } {\sum\limits_j  {A_{i j } \hat \psi ''_j  } } \nonumber \\
q_0 &=& \sum\limits_{i  \in R_2 } {\hat \psi '_i  } \sum\limits_{i  \in R_1 } {\sum\limits_j  {A_{i j } \hat \psi '_j  } }  - \sum\limits_{i  \in R_1 } {\hat \psi '_i  } \sum\limits_{i  \in R_2 } {\sum\limits_j  {A_{i j } \hat \psi '_j  } }
\label{eq:coefficients}
\end{eqnarray}
The algorithm is to apply $A$ repeatedly until two real solutions $\eta_1$ and $\eta_2$ for (\ref{eq:quadratic}) exist.  One solution will be then used to guide further iterations to $(\lambda_1,\psi_1)$; the other, to $(\lambda_2,\psi_2)$. 

The power method becomes \cite{gubernatis}: choose two starting points $\psi'$ and $\psi''$, which need not be orthogonal, then for each iteration step, compute
\begin{eqnarray}
\psi' &\leftarrow& \psi' / \|\psi'\| \nonumber \\
\psi'' &\leftarrow& \psi'' / \|\psi''\| 
\end{eqnarray}
and if the roots of (\ref{eq:quadratic}) are real \cite{complex}, update using 
\begin{eqnarray}
\psi'  &\leftarrow& A \psi'' + \eta_1 A \psi' \nonumber \\
\psi'' &\leftarrow& A \psi'' + \eta_2 A \psi'
\label{eq:update}
\end{eqnarray}
otherwise use
\begin{eqnarray}
\psi' &\leftarrow& A \psi' \nonumber \\
\hat\psi'' &\leftarrow& A \hat\psi''
\end{eqnarray}
Eigenvalues can be estimated from
\begin{eqnarray}
\lambda_1  &=& \frac{ \sum\limits_{i\in R_1}\sum\limits_j A_{ij}\psi_j'
              +\eta_1 \sum\limits_{i\in R_1}\sum\limits_j A_{ij}\psi_j'' }
                    { \sum\limits_{i\in R_1}\psi_i'
               +\eta_1\sum\limits_{i\in R_1}\psi_i'' }\nonumber \\
\lambda_2  &=& \frac{ \sum\limits_{i\in R_1}\sum\limits_j A_{ij}\psi_j'
              +\eta_2 \sum\limits_{i\in R_1}\sum\limits_j A_{ij}\psi_j'' }
                    { \sum\limits_{i\in R_1}\psi_i'
               +\eta_2\sum\limits_{i\in R_1}\psi_i'' }
\label{eq:eigenvalues}
\end{eqnarray}
where $\eta_1$ and $\eta_2$ generate the largest and next largest eigenvalue estimates.

\section{Monte Carlo Implementation}

The basic operation of a power method is a matrix-vector multiplication. Here, we now describe how we used the Monte Carlo method to estimate the repetition of such multiplications.  

In the basis defining the matrix elements of $A$, we write
\begin{eqnarray}
 \psi'  &=& \sum\limits_i {\omega_i ' } \left| i \right\rangle  \nonumber \\ 
 \psi'' &=& \sum\limits_i {\omega_i ''} \left| i \right\rangle 
 \end{eqnarray}
We will call the amplitudes $\omega'_i$ and $\omega''_i$  {\it weights\/} even though they are not necessarily all positive nor are the sums of their absolute values unity. We will also assume that the elements $A_{ij}$ of the $M\times M$ matrix $A$ are easily generated, as we are ultimately interested in cases where $M$ is so large that this matrix must be generated on-the-fly as opposed to being stored. Next, we imagine we have $N$ particles distributed over the $M$ basis states defining $A$. Generally, $N\ll M$.  Then, at each iteration step, we interpret $A_{ij}$ as the weight of a particle arriving in state $|i\rangle $ per unit weight of a particle in state $|j\rangle $. The action of $A_{ij}$ on a $\psi$ thus causes all particles currently in state $|j\rangle$ to jump to $|i\rangle$, carrying to $|i\rangle$ their current weight $\omega_j$, modified by $A_{ij}$.  

The jumps will be executed probabilistically. To do this we let the total weight leaving state $|j\rangle$ be
\begin{equation}
W_j=\sum_i A_{ij}
 \label{eq:weight_multiplier}
\end{equation}
and define the transition probability from $|j\rangle$ to $|i\rangle$ be
\begin{equation}
T_{ij}=A_{ij}/W_j
 \label{eq:transition_probability}
\end{equation}
The number $W_j$ is called the state weight multiplier. How we use these densities will depend on the size and types of the matrices under consideration.

\subsection{Medium Matrices}

If $M$ is  sufficiently small so we can store all components of our vectors, then a Monte Carlo procedure for jumping is easily constructed. Instead of always (i.e., with probability 1) moving weight $A_{ij}$ from state $|j\rangle$ to state $|i\rangle$, we will instead sample a $|i\rangle$ from $T_{ij}$ and multiply the transferred weight by the ratio of the true probability (1.0) to the sampled probability ($T_{ij}$); that is, if state $|i\rangle$ is sampled, the weight arriving in state $|i\rangle$ from $|j\rangle$ is multiplied by
 \begin{equation}
A_{ij} \frac{1.0}{T_{ij}}=A_{ij} \frac{1.0}{A_{ij}/W_j}=W_j
\end{equation}

As for many Monte Carlo simulations, as is the case for the transfer matrix of the Ising model, the particle weights defining the eigenvector associated with the largest eigenvalue can be made all positive. The second eigenfunction however must be represented by some particles of negative weight and some particles of positive weight. These negative and positive weights must for some jumps at least partially cancel to maintain a correct estimation of the second eigenfunction. When $N\ll M$, this cancellation does not occur often enough in a Monte Carlo simulation without proper design: because the number of states vastly out numbers the number of particles, the probability that a negatively and a positively weighted particle randomly arrive in the same state becomes trivially small.
 
There are several ways to arrange the cancellation \cite{booth1}. We found the Arnow {\it et al.\/} \cite{arnow} algorithm  effective and convenient.  First, we consider two particles of weights $w_1$ and $w_2$ in states $|j_1\rangle$ and $|j_2\rangle$, and then let $T_{i j_1 }$ and $T_{i j_2 }$ be the probabilities of their reaching state $|i\rangle$. Next we use  the weight multiplier $W_j$ for jumping from state $|j\rangle$ to state $|i\rangle$ and suppose that states $|i_1\rangle$ and $|i_2\rangle$ are sampled from the density $T_{i  j_1}+T_{i  j_2}$. This density can be sampled by sampling a $|i_1\rangle$ from $T_{i  j_1}$ and a $|i_2\rangle$ from $T_{i  j_2}$. The true probability that particle 1 jumps to $|i_1\rangle$ is $T_{i_1  j_1}$ and the true probability that particle 2 jumps to $|i_1\rangle$ is $T_{i_1  j_2}$. The ratio of the true density to the sampled density for particle 1 is
\begin{equation}
 \frac{T_{i_1  j_1}}{T_{i_1  j_1}+T_{i_1 j_2}}
\end{equation}
so that the weight arriving at state $|i_1\rangle$ from particle 1 is
\begin{equation}
 \frac{w_1W_{j_1} T_{i_1  j_1}}
 {T_{i_1  j_1}+T_{i_1 j_2}}
\end{equation}
The true probability that particle 2 arrives at $|i_1\rangle$ is $T_{i_1 j_2}$. The ratio  of the true density to the sampled density for particle 2 is
\begin{equation}
 \frac{T_{i_1 j_2}}
  {T_{i_1  j_1}+T_{i_1 j_2}}
\end{equation}
so that the weight arriving at $|i_1\rangle$ from particle 2 is
\begin{equation}
 \frac{w_2 W_{j_2} T_{i_1 j_2}}
   {T_{i_1  j_1}+T_{i_1 j_2}}
\end{equation}
Thus the total weight arriving at $|i_1\rangle$ is
\begin{equation}
 \frac{
 w_1W_{ j_1} T_{i_1  j_1}
 + w_2 W_{j_2} T_{i_1  j_2}}
  {T_{i_1  j_1}+T_{i_1  j_2}}
\end{equation}
By similar arguments, the total weight arriving at $|i_2\rangle$ is
 \begin{equation}
 \frac{
 w_1W_{j_1} T_{i_2  j_1}
 + w_2 W_{j_2} T_{i_2  j_2}}
    {T_{i_2  j_1}+T_{i_2  j_2}}
\label{eq:the_above}
\end{equation}
 
We note that to get meaningful cancellation, say between particle 1 with weight $w_1<0$ and particle 2 with weight $w_2>0$, the transition probabilities must overlap somewhat. For example, if in (\ref{eq:the_above}) $T_{i_2  j_1}=\epsilon \ll 1$, then the total weight arriving at $i_2$ is essentially just the weight arriving of particle 2 alone.
\begin{equation}
 \frac{
 w_1W_{j_1} \epsilon
 + w_2 W_{j_2} T_{i_2  j_2}}
 {\epsilon+T_{i_2  j_2}}
 \approx
  w_2 W_{j_2}
 \end{equation}
How one arranges better overlap is problem dependent. For our Ising simulations, we sorted the particles into state order (a state is represented by the bits of an integer). Particles 1 and 2 were then sampled together according to the Arnow {\it et al.\/} scheme, then particles 3 and 4 are sampled together, and so forth. The fact that the list is ordered means that there are (typically) many nearby states $|i\rangle$ that are accessible from both particles $\ell$ and $\ell+1$ with nontrivial transition probabilities.

As the iteration progresses, the absolute value of the weights of some particles becomes very large, and those of some others, very small. As standard for Monte Carlo methods with weighted particles, particles with weights of small magnitude are stochastically eliminated and those with large magnitudes are stochastically split. To do this we used a procedure called the comb \cite{comb1,comb2}. It is described in Appendix~A. 

The steps of the algorithm are: First, we initialize the states and weights of two vectors. For the Ising simulation, each vector had the same states but different weights. We selected the states uniformly and randomly over the interval $(0,2^m-1)$ and selected the $\omega_i'$ uniformly and randomly over the interval (0,1) and the $\omega_i''$ uniformly and randomly over the interval (-0.5,0.5). Then, for a fixed number of times we iterate. At each iteration we execute the jump procedure for each particle, place the particle list in state order, effect cancellations, estimate the eigenvalues from (\ref{eq:eigenvalues}), update $\psi'$ and $\psi''$, and then comb.   $R_1$ consisted of the states for which more than half of its $m$ bits were $0$'s and $R_2$ consisted of the states for which more than half of its $m$ bits were $1$'s. Additional algorithmic details  are given in Appendix~B.

Table \ref{table:1} shows the computed $\lambda_1$ and $\lambda_2$ for $m=12$.  A computer program was written to make 20 independent (different random number seeds) calculations using $N$ particles per iteration, for various values of $N$.  Note that even when $N=100<<M=2^{12}=4096$ the method still separates the eigenfunctions. The simulations were run at the critical temperature of the infinite lattice; that is, $\nu=0.4406867935097715$ \cite{thompson}. The last line in the table gives the eigenvalues deterministically obtained by using (\ref{eq:exact}). From the last few lines, we also see that when the number of particles floods the number of states, that is, when all the basis states are being used multiple times, exceptional accuracy is obtained. 

Also in the table is the timing for each run. The runs were done on a single 1.5 GHz Mac PowerPC processor. We note that the runs appear to scale sub-linearly with the number of particles. This scaling is deceiving. For a small number of particles, the time required to set up the transition matrix $A_{ij}$ is significant compared to the rest of the calculation. The {\it Monte Carlo} part is dominated by the state order sort which scales as $N\log{N}$, so as $N$ becomes very large, the run time scaling should eventually be slightly super-linear. 

\begin{table}
\caption{Particle number $N$ dependence of individual runs for the $m=12$ Ising Model.  The $\lambda_i$ and $\sigma_i$ ($i=1,2$) are run's eigenvalue averages and error estimates.  The row labeled ``Onsager'' are the eigenvalues obtained from (\ref{eq:exact}).  Also given is the wall clock time for each run. The order of the transfer matrix is 4096. \label{table:1}}
\begin{ruledtabular}
\begin{tabular}{lccccr}
  N & $\lambda_1$ & $\sigma_1$&    $\lambda_2$&   $\sigma_2$ &minutes \\
  \noalign{\smallskip}
  \hline
 \noalign{\smallskip}
        100&$71415.164$ &$65$   & 67021.909&103   & 5.4\\
      1000&$71527.110$ &$17$   & 67023.420&  31    & 7.9\\
    10000&$71553.325$ &$5.3$  & 66956.314&   9.1  & 22.8\\
   100000&$71557.854$ &$2.0$  & 67005.486&  3.2   & 150.0\\
 1000000&$71557.129$ &$0.36$& 67010.989&  0.58 & 1474.0\\
\hline
 Onsager&$71557.047$ & & $67010.869$ & &  \\
\end{tabular}
\end{ruledtabular}
\end{table}

\subsection{Large Matrices}
 
For $M\le 2^{12}$, sampling from the cumulative probability $C_i=\sum_{k=0}^i T_{kj}$ works well. If the number of states gets too large ($M=2^{12}$ was our limit), then $C_i$ cannot be sampled directly because it cannot fit in the computer's memory.  In this case, we could just randomly pick from any state $|j\rangle$ any state $|i\rangle$ with probability $1/M$ instead of always picking a state $|i\rangle$ (i.e., picking it with probability 1). Then, if state $|i\rangle$ is sampled, the weight of particles arriving in state $|i\rangle$ is
 \begin{equation}
A_{ij} \frac{1.0}{1/M}=A_{ij} M
\end{equation}
The problem with this approach is that the $A_{ij}$ can have immense variation so that this simple sampling scheme is unlikely to work well as a Monte Carlo method. This situation is especially true for the Ising problem. A large part of such variations however can be removed by sampling the new state in stages and then sewing the stages together. 

To explain the sewing procedure \cite{booth08}, we will first assume that we can write any state $|i\rangle$ in our basis as a direct product of the states in a smaller basis, for example, 
\[
\left| i \right\rangle  = \left| {i_2 } \right\rangle \left| {i_1 } \right\rangle 
\]
Instead of transferring weight
\begin{equation}
W_j=\sum_k A_{kj}
\end{equation}
from state $|j\rangle$ to state $|i\rangle$ with probability
\begin{equation}
T_{ij}=A_{ij}/\sum_k A_{kj}=A_{ij}/W_j,
\end{equation}
we will use the $a_{ij}$ that would apply to the smaller set of states and then make an appropriate weight correction.

For each smaller set of states, we rewrite the analogous transition probability from state $|j\rangle$ to state $|i\rangle$ as
\begin{equation}
t_{ij}=a_{ij}/w_j,
\label{eq:trans_mtrx}
\end{equation}
and the analogous weight multiplier as
\begin{equation}
w_j=\sum_k a_{kj}
\label{eq:wgt_mult}
\end{equation}
We thus will sample $|i_1\rangle$ and $|i_2\rangle$ from the probability function
\begin{equation}
  t_{i_2j_2}   t_{i_1j_1}
\label{eq:trans_prob}
\end{equation}
Now, we define $C_{ij}$ to be the weight correction necessary to preserve the expected weight transfer from state $|j\rangle=|j_2\rangle|j_1\rangle$ to state $|i\rangle$. It satisfies
\begin{equation}
A_{ij}=  C_{ij}  t_{i_2j_2}   t_{i_1j_1}
\label{eq:transfer_mtrx}
\end{equation}
Using (\ref{eq:trans_mtrx})  in (\ref{eq:transfer_mtrx}), we find
\begin{equation}
A_{ij}=  C_{ij}  \frac{a_{i_2j_2}}{w_{j_2} }
                         \frac{a_{i_1j_1}}{w_{j_1} }
\label{eq:mtrx}
\end{equation}
Thus
\begin{equation}
C_{ij} = w_{j_1} w_{j_2} \frac{A_{ij}} {a_{i_1j_1} a_{i_2j_2} }
\label{eq:weight_correction}
\end{equation}

This sewing method generalizes easily. For $k$ sets of states, (\ref{eq:mtrx})  and (\ref{eq:weight_correction})  become 
\begin{equation}
A_{ij}=  C_{ij}  \prod_{n=1}^k t_{i_nj_n}
\end{equation}
and
\begin{equation}
C_{ij}=A_{ij} \prod_{n=1}^k\frac{w_{j_n}}{a_{i_nj_n}}
\end{equation}
For the Ising problem, the weight correction is 
\begin{equation}
C_{ij} =\exp(\nu D_i)\prod_{n=1}^k w_{j_n}
\end{equation}
where $\nu D_i$ is the energy difference per $k_BT$ between calculating with the bits together and the bits separately \cite{booth08}.


\begin{table*}[bht]
\caption{Estimates of the two dominant eigenvalues $\lambda_1$ and $\lambda_2$ and their errors of the transfer matrix for variously-sized two-dimensional Ising models. Each estimate was based on 20 independent simulations. Also given for each lattice size is value for $\lambda_1$ computed via Onsager's exact result \cite{thompson,onsager}. \label{table:2}}
\begin{ruledtabular}
\begin{tabular}{lllll}
  Matrix Size & $\lambda_1$ (Onsager)& $\lambda_1$ & $\lambda_1$ (Onsager)& $\lambda_2$\\
 \hline
 $2^{16}\times 2^{16}$  & $2.93297\times 10^6$       & $2.93307 \pm 0.00008\times 10^6$
                                      & $2.79225\times 10^6$       & $2.79482 \pm 0.00010\times 10^6$\\
 $2^{24}\times 2^{24}$  & $4.95473\times 10^9$       & $4.95480 \pm 0.00020\times 10^9$ 
                                      & $4.79510\times 10^9$       & $4.79502 \pm 0.00029\times 10^9$\\                                                                             
 $2^{32}\times 2^{32}$  & $8.39316\times 10^{12}$   & $8.39311\pm  0.00049\times 10^{12}$
                                      & $8.18959\times 10^{13}$   & $8.18807\pm  0.00061\times  10^{12}$\\
 $2^{40}\times 2^{40}$  & $1.42333\times 10^{16}$   & $1.42334\pm  0.00007\times  10^{16}$ 
                                      & $1.39565\times 10^{17}$   & $1.39558\pm  0.00008\times  10^{16}$ \\                                                                                       
 $2^{48}\times 2^{48}$  & $2.41504\times 10^{19}$   & $2.41522 \pm 0.00019\times 10^{19}$
                                      & $2.37584\times 10^{19}$   & $2.37481 \pm 0.00054\times 10^{19}$ \\
\end{tabular}
\end{ruledtabular}
\end{table*}

Using this sewing algorithm for the sampling of states, we computed the first and second eigenvalues and their standard deviations for variously sized two-dimensional Ising models by sewing sets of 8 bits. The results are shown in Table \ref{table:2}. Twenty independent runs were done for each size with 500 iterations per run. There were 1 million particles per iteration for $m=16,24$, and 32 and 5 million for $m=40$ and 48.  Only the second half of the iterations for each run used in the estimation process. This choice for ``burn-in" was arbitrary and excessive. Typically the iteration converges to its fixed point in about 10 or fewer steps. Clearly, the estimated mean of the eigenvalues are consistent with the analytic result of Onsager.  The $m=48$ run took about 2.15 hours on a single 1.25 GHz Alpha EV6 processor.  $\nu=0.4406867935097715$, the bulk critical value. $R_1$ consisted of the states for which more than half of its $m$ bits were $0$'s and $R_2$ consisted of the states for which more than half of its $m$ bits were $1$'s.  In Table~\ref{table:3}, we present the run averages for the $M=2^{48}$ case.  We note that $2^{48}\approx 2.8\times 10^{14}$.


\begin{table}
\caption{Run dependence of $\lambda_1$ and $\lambda_2$ for the  $48\times 48$ Ising Model. Each run had 5 million particles and 500 iterations. Only the last 250 were used to computed the stated run averages.  \label{table:3}}
\begin{ruledtabular}
\begin{tabular}{lll}
 Run& $\lambda_1$ &$\lambda_2$\\
 \hline
  1& $2.415237261676499\times 10^{19}$ &$2.376562826996199\times 10^{19}$ \\
  2& $2.416053055304024\times 10^{19}$ & $2.376357592801913\times 10^{19}$ \\
  3& $2.414208101931427\times 10^{19}$ & $2.373861137045189\times 10^{19}$ \\
  4& $2.414240257597889\times 10^{19}$ & $2.374603969362707\times 10^{19}$ \\
  5& $2.415129736468956\times 10^{19}$ & $2.375668969910789\times 10^{19}$\\
  6& $2.414540994327886\times 10^{19}$ &$2.367761547315434\times 10^{19}$ \\
  7& $2.415931928295190\times 10^{19}$ & $2.371678054324882\times 10^{19}$ \\
  8& $2.416760807133448\times 10^{19}$ & $2.380200781153834\times 10^{19}$ \\
  9& $2.413931047178054\times 10^{19}$  & $2.374313310822579\times 10^{19}$ \\
 10& $2.414288017604171\times 10^{19}$ &  $2.374183202851979\times 10^{19}$ \\
 11& $2.415907427261585\times 10^{19}$ &  $2.374280853036757\times 10^{19}$ \\
 12& $2.415718175365685\times 10^{19}$ & $2.374828531467586\times 10^{19}$ \\
 13&$2.415613030894034\times 10^{19}$  &  $2.374305559747647\times 10^{19}$ \\
 14& $2.414272502002316\times 10^{19}$ & $2.374687824232260\times 10^{19}$ \\
 15& $2.416636076568460\times 10^{19}$ & $2.376864186656784\times 10^{19}$ \\
 16& $2.414631286368427\times 10^{19}$ &  $2.376097488856122\times 10^{19}$ \\
 17& $2.415202515765169\times 10^{19}$ & $2.376875902411582\times 10^{19}$ \\
 18& $2.415187019163818\times 10^{19}$ & $2.373320142080186\times 10^{19}$ \\
 19& $2.414692528984318\times 10^{19}$ & $2.374355064690523\times 10^{19}$ \\
 20& $2.416154446732387\times 10^{19}$ & $2.375485037469832\times 10^{19}$ \\
\end{tabular}
\end{ruledtabular}
\end{table}

\section{Concluding Remarks}
 
We have presented a Monte Carlo algorithm that enables the determination of two extremal eigenvalues of a very large matrix. The explicit demonstration of the power of our algorithm was the determination of the two largest eigenvalues of the transfer matrix of the two-dimensional Ising model in a zero magnetic field. The convenience of this matrix was the existence of exact expressions for its eigenvalues for any finite-sized system.  We were able to reproduce the exact values for the two dominant eigenvalues to within the statistical error of our simulations. 

Physics and chemistry provide numerous problems where obtaining several extremal eigenvalues is important. One problem class is finding the ground state and a few excited states of atoms, molecules, solids, and nuclei \cite{note}. For these quantum problems, quantum Monte Carlo Monte Carlo (QMC) projector methods \cite{hammond,zhang} are commonly used. In fact, it was for these methods that the Arnow {\it et al\/}. \cite{arnow} method used here was first proposed as a method for taming the fermion sign problem  \cite{hammond}.  We comment that our new algorithm requires sampling from the second eigenfuction which must have positive and negative components just as a fermion state must also have. Our successful cancellation of signed particles representing this eigenfunction and the slightly superlinear scaling of our computation time with system complexity, as opposed to an exponential scaling,  demostrates that we have ``solved" the sign problem \cite{sign} for our application.

We also note that several investigators \cite{nightingale} have adapted the technology of the diffusion QMC method \cite{hammond} to find the dominant eigenvalue of the transfer matrix of various classical spin models. This Monte Carlo approach is distinctly different from ours even if ours were restricted  to just the dominant state. For example, we do not use an importance function to guide the sampling nor do we use back propagation to enhance estimators. The current extension of diffusion QMC to the concurrent calculation of multiple eigenpairs \cite{hammond,nightingale3} is also quite different from our extension. Diffusion Monte Carlo analogs the simultaneous iteration method mentioned in Section III. (Still another approach to concurrent eigenvalue estimation was used by Hasenbusch {\it et al\/}.\cite{hasenbusch}.) We further note Nightingale and Bl\"ote's  \cite{nightingale2} use of a Monte Carlo power method to find the second eigenvalue of a Markov chain transition matrix satisfying detailed balance. For this case the dominant eigenpair is known {\it a priori} and was used to deflate the matrix so the projection was to the second largest eigenvalue.  We comment that Booth \cite{booth1} has presented an algorithm for determining a single  eigenvalue without the need to know or concurrently determine any other one.

Our transfer matrix was positive, asymmetric, and dense. How is our algorithm changed if the matrix lacks one or more of these properties? For simple test cases, we have successfully constructed deterministic procedures for matrices whose elements are complex valued. Also for simple test cases, we have had success for real asymmetric matrices whose eigenvalues are complex valued. Devising Monte Carlo algorithms for real, symmetric, sparse matrices has however received more of our attention \cite{gubernatis2}.  

For an indefinite but symmetric matrix, the symmetry and reality assure the reality of the eigenpairs, but do not fix the sign of the dominant eigenvalue or the components of its eigenvector. Such matrices can also have the dominant eigenvalue being degenerate.  In our algorithm a signature of the dominant eigenvalue being degenerate is the quantity $q_1$ (\ref{eq:coefficients}) in our quadratic equation (\ref{eq:quadratic}) becoming extremely small. In cases where we knew the degeneracy {\it a priori\/}, we saw our estimate of $\lambda_1$ being very close to $\lambda_2$, that is, we saw our estimates of the dominant degenerate eigenvalue being slightly but artificially separated. This type of difficulty is inherent to a power method. If the matrix elements are of mixed sign, the modifications to the algorithm are quite simple: we take the absolute value of the $A_{ij}$ and multiply the weight multiplier  by the sign of $A_{ij}$. We also have to account for the sign of the weights $\omega_i'$  and $\omega_i''$ appropriately. Weight cancellation must be done with care.  If the matrix is sparse, other options for selecting the state to jump to become available that avoid storing the cumulative distribution, improve efficiency, and reduce variance \cite{gubernatis2}. 

Here, we have focused principally on the determination of  eigenvalues, noting the rapid convergence to their values, but what about the determination of the eigenvectors?  If we can store the vectors in computer memory we can accurately determine them.  For a problem in a continuum, for example, finding the dominant eigenvalues of an integral equation such as those occurring for the transport of neutrons, we have accurately estimated the two dominant eigenfunctions with an efficiency enhanced over the basic power method for determining just the dominant eigenfunction \cite{booth3}.

The dynamics of determining eigenvectors is as follows: Because the standard power method for the dominant eigenpair converges to $\psi_1$ as $\lambda_2/\lambda_1$ (\ref{eq:pm}),  the dominant eigenvector becomes increasingly difficult to determine as $\lambda_1$ and $\lambda_2$ become very close in magnitude.  Because we are projecting to the first two dominant eigenpairs, our method converges to $\psi_1$ as $\lambda_3/\lambda_1$. This is usually a gain relative to the standard power method. Convergence to $\psi_2$ goes as $\lambda_3/\lambda_2$. This ratio illustrates the fact that in principle we can accelerate convergence to $\psi_i$ by seeking the $L>i$ highest eigenfunctions and thus converging to $\psi_i$ as $\lambda_{L+1}/\lambda_i$. To do this, we need to use $L$ groupings and $L$ starting states and adjust the sum of the iterated states to guide them to convergence to $L$ different eigenvalues. We have observed accelerated convergence for the $L=3$ case on simple test cases. We comment that finding the eigenvectors of the transfer matrix of the Ising and other models is a problem for which finding the eigenvectors becomes increasingly difficult as the system size increases, if we could in fact store them. For this type of matrix, the eigenvalue spectra is expected to become quasi-continuous starting with the second eigenvalue and the first and second eigenvalues become asymptotically degenerate at a critical point  \cite{privman}.

In closing, we believe that the algorithm presented here is accurate, easy to implement, and applicable to many other problems. Different problems afford the opportunity to improve its efficiency by modifying some of its details. Wider use of the algorithm will define more crisply its strengths and limitations than is possible by just the present application.


\appendix
\section{}

The comb is a stochastic procedure for selecting $M$ particles, not necessarily all different, from a list of $N$ unevenly weighted particles and preserving the total weight.  If $\omega_i>0$ is the weight of an individual particle in the original list, its weight in the new list becomes $W_T/M$  where $W_T=\sum_i^N \omega_i$ is the total weight. To effect this procedure, we construct the cumulative sums of the original weights
\begin{equation}
C_j  = \sum\limits_{i = 1}^j {\omega_i } 
\end{equation}
where $j=0,1,2,\dots,N$ and $C_0=0$ and $C_N=W_T$, draw a random number $\xi$ uniformly distributed over (0,1), and then for $k=1,2,\dots,M$, select particle $j$ for the new list if
\begin{equation}
C_{j-1}  < (k-1+\xi)\frac{{W_T }}{M} \le C_j 
\end{equation}
Dependent of the size of the difference $C_{j-1}-C_j$ relative to $W_T/M$, particle $j$ is selected zero, one, or more times. The procedure calls the random number generator only once, and the new particle list is always of a predetermined size.

In the present simulations we have a list of particles and to each particle $i$ is associated a state $|i\rangle$ and two weights, $\omega_i'$ and $\omega_i''$. If we were to comb the two lists of weights separately, we would in general produce two lists of states. We wanted one such list. Accordingly, we combed in the following manner: we formed
\begin{eqnarray}
 p_i '   &=& \frac{{|\omega_i ' |}}{{\sum\limits_{i = 1}^N {|\omega_i ' | } }} \\
 p_i ''   &=& \frac{{|\omega_i '' | }}{{\sum\limits_{i = 1}^N {|\omega_i'' | } }}
 \end{eqnarray}
and then generated the cumulative distribution.
\begin{equation}
C_j  = \sum\limits_{i = 1}^j \frac{1}{2}(p_i '+p_i'') 
\end{equation}
Now we combed the $C_j$ as before, noting that we have chosen a normalization such that $W_T=1$. Instead of giving the selected particle $j$ a weight $W_T/M$ for both $w_j'$ and $w_j''$, we instead assign 
\begin{eqnarray}
 \omega_j'   &=&  \frac{{p_i '  {\rm sign}(\omega_i')}}{{p_i '  + p_i '' }} \\
 \omega_j''  &=&  \frac{{p_i'' {\rm sign}(\omega_i'')}}{{p_i '  + p_i '' }} 
\end{eqnarray}

\section{}

The following are some useful particulars of our Monte Carlo implementation of our modified power method. First, instead of updating via (\ref{eq:update}) we update via
\begin{eqnarray}
\psi'  &\leftarrow& A \psi' + \eta_1 A \psi'' \nonumber \\
\psi'' &\leftarrow& \frac{1}{\eta_2}A \psi' +  A \psi''
\end{eqnarray}
This form is more symmetric and makes more explicit that we are trying to remove $\psi''$ from $\psi'$ and vice versa. To avoid overcorrecting we have sometimes found it useful to update via
\begin{eqnarray}
\psi'  &\leftarrow& A \psi' + \eta'_1 A \psi'' \nonumber \\
\psi'' &\leftarrow& \frac{1}{\eta_2'}A \psi' +  A \psi''
\end{eqnarray}
where
\begin{eqnarray}
\eta'_1&=&{\rm sign}(\eta_1)\min(\alpha,|\eta_1|)\\
\eta'_2&=&{\rm sign}(\eta_2)\min(\alpha,|\eta_2|)
\end{eqnarray}
and $\alpha$ is some small positive number.

Besides the eigenvalue estimators (\ref{eq:eigenvalues}), there are others (\ref{eq:groupings}) equally valid that provide multiple cross-checks to (\ref{eq:eigenvalues}). These include
\begin{eqnarray}
\lambda_1  &=& \frac{ \sum\limits_{i\in R_1}\sum\limits_j A_{ij}\psi_j'}
                                  { \sum\limits_{i\in R_1}\psi_i'}
                      =   \frac{ \sum\limits_{i\in R_2}\sum\limits_j A_{ij}\psi_j' }
                     { \sum\limits_{i\in R_2}\psi_i'} \\
\lambda_2  &=& \frac{ \sum\limits_{i\in R_1}\sum\limits_j A_{ij}\psi_j''}
                                 { \sum\limits_{i\in R_1}\psi_i''}=
                           \frac{\sum\limits_{i\in R_2}\sum\limits_j A_{ij}\psi_j'' }
                                  { \sum\limits_{i\in R_2}\psi_i''}
\end{eqnarray}
all of which involve sums needed for (\ref{eq:eigenvalues}) so they cost nothing extra to compute. Another use of these estimators is monitoring the utility of the regions $R_1$ and $R_2$, as an estimator will perform inconsistently if too few particles occupy a region or if the sum of the weights in a given region is small. The most useful regions are those that are a major subset of the positive and negative regions of the second eigenvector.

For the Ising model we obtained some additional useful weight cancellation by exploiting the fact that most of the particles are in fully magnetized states and before using the comb replacing all the particles in the ``up" state by a single particle whose weight is the sum of the ``up" weights of these particles and doing similarly for the particles in the ``down" state.

Because the $\omega'_i$ should be positive, after combing we explicitly set each $\omega'_i$ equal to its absolute value. THe step is useful because even though we initialize $\psi'$ to have all positive components, the weight cancelation procedure can make some of them negative. Once so, they might project out rapidly and may adversely affect the eigenpair estimation until they do.

\begin{acknowledgments}
We thank M. E. Fisher for a helpful conversation. We gratefully acknowledge support of the U. S. Department of Energy through the LANL/LDRD program.
\end{acknowledgments}
 

\begin{thebibliography}{}
\bibitem{complex}
If complex arithmetic is used, the need to wait for the reality of these roots can be obviated.
\bibitem{note}  if the magnitude of the ground state eigenvalue is smaller than that of the dominant eigenvalue, convergence to the ground state  is reasonably accomplished by appropriately shifting the eigenvalue spectrum, $A\rightarrow A-\sigma I$ \cite{wilkinson}.
\end{thebibliography}

\end{document}